# Crown ether decorated silicon photonics for safeguarding against lead poisoning

Luigi Ranno[1,†], Yong Zen Tan[2,†], Chi Siang Ong[2], Xin Guo[3], Khong Nee Koo[4], Xiang Li[3], Wanjun Wang[3], Samuel Serna[1], Chongyang Liu[5], Rusli[3], Callum G. Littlejohns[6], Graham T. Reed[6], Juejun Hu[1], Hong Wang[3] and Jia Xu Brian Sia[1,3,*]

[1]*Department of Materials Science & Engineering, Massachusetts Institute of Technology, Cambridge, M.A., USA*

[2]*Fingate Technologies, Singapore*

[3]*School of Electrical and Electronic Engineering, Nanyang Technological University, 50 Nanyang Avenue, 639798, Singapore*

[4]*Vulcan Photonics, Kuala Lumpur, Malaysia*

[5]*Temasek Laboratories, Nanyang Technological University, 50 Nanyang Avenue, 637553, Singapore*

[6]*Optoelectronics Research Centre, University of Southampton, Southampton SO17 1BJ, UK*

[†]*These authors contributed equally*

*\*Corresponding author: jxbsia@mit.edu, jiaxubrian.sia@ntu.edu.sg*

**Lead ($Pb^{2+}$) toxification in society is one of the most concerning public health crises that remains unaddressed. The exposure to $Pb^{2+}$ poisoning leads to a multitude of enduring health issues, even at the part-per-billion scale (ppb). Yet, public action dwarfs its impact. $Pb^{2+}$ poisoning is estimated to account for 1 millions deaths per year globally, which is in addition to its chronic impact on children. With their ring-shaped cavities, crown ethers are uniquely capable of selectively binding to specific ions. In this work, for the first time, the synergistic integration of highly-scalable silicon photonics, with crown ether amine conjugation via Fischer esterification in an environmentally-friendly fashion is demonstrated. This realizes a photonic platform that enables the *in-situ*, highly-selective and quantitative detection of various ions. The development dispels the existing notion that Fischer esterification is restricted to organic compounds, laying the ground for subsequent amine conjugation for various crown ethers. In this work, the platform is engineered for $Pb^{2+}$ detection, demonstrating a large dynamic detection range of 1–262000 ppb with high selectivity against a wide range of relevant ions. These results indicate the potential for the pervasive implementation of the technology to safeguard against ubiquitous lead poisoning in our society.**

# 1. Introduction

Anthropogenic lead poisoning represents one of the primary public health concerns since antiquity[1]. $Pb^{2+}$ is a cumulative toxicant that leads to multi-faceted impact on biological functions in the long term [2–11]. $Pb^{2+}$ has the affinity to substitute other bivalent and monovalent cations. For instance, $Pb^{2+}$ can replace $Ca^{2+}$ ions to cross the blood-brain barrier, resulting in neurological deficits[7]. This effect is exacerbated in children due to the ongoing development of their neurological and nervous system[8]. $Pb^{2+}$ is also found to impact cardiac function, causing reduction in the speed of heart contraction and relaxation[10]. Furthermore, fetal exposure can result in a wide array of risks during pregnancy[9]. The above examples only serve to highlight a non-exhaustive overview of the impact of lead on our society, and many other detailed studies are available to interested readers[11–16]. However, public action against lead toxification is disproportional to its impact. It has been estimated that lead service lines still deliver drinking water to about ten million households in the US alone [17]. The impact of lead leads to the common conclusion that there should be zero-tolerance to lead exposure[18]. To that effect, the Environmental and Energy Law Program (EPA), US has implemented a limit of 15 parts-per-billion (ppb) in drinking water[19]. Lead poisoning is even more pronounced in developing countries, where the World Health Organization (WHO) estimates that of the 240 million people that are overexposed, 99 % comes from developing countries [20,21]. Lead exposure accounts for more than one million deaths annually, with significant societal and economic costs, specifically in developing countries[22]. These facts highlight the urgency for the development of technologies that guards against lead toxification.

Contemporary methods for lead detection can be grouped into two primary categories: Inductively Coupled Plasma-Mass Spectrometry/Optical Emission Spectroscopy (ICP-MS/OES)[23,24], and colorimetric test strips[25]. The former represents the state-of-art, but however, suffers from low sample throughput, requiring lengthy and expensive sample preparation and analysis by trained personnel. Furthermore, such systems are dedicated for lab use only, and not viable for on-site, *in-situ* analysis[24]. Colorimetric test strips, while low-cost and widely available, are qualitative and might lack accuracy in detection[25]. The development of $Pb^{2+}$ sensors presented in this work combines the advantages of both technologies: highly quantitative and selective sensing as well as rapid, portable detection capabilities.

The photonic sensor platform makes use of crown ethers, cyclic polyethers consisting of multiple oxygen atoms forming a ring structure[26,27]. This class of compounds were first synthesized by Charles Pederson in the 1960s, who was subsequently awarded the 1967 Nobel prize for this discovery[28,29]. As a result of the cavity which arises from the ring structure, crown ethers possess a remarkable ability to selectively bind to certain ions or molecules based on their properties such as size selectivity charge accommodation, ring geometry and structure energetic favorability[28–30]. Thus far, crown ethers have employed electrochemical[31–33] and fluorescent[34–36]-based detection schemes. However, scaling these technologies to low-cost large sensor arrays for widespread detection remains a major challenge.

In this manuscript, we demonstrated, for the first time, a crown ether functionalized silicon photonics platform. Traditionally, the functionalization of crown ether on silicon involves the use of silylating agents with trisubstituted silyl groups which are moisture and pH sensitive, and require stringent process control[37,38]. In the above protocol, the reagents can potentially undergo self-reaction, resulting in agglomeration, which decreases surface uniformity and negatively impacts sensor reproducibility[37,38]. The application of the Fischer esterification protocol[39] to couple carboxylic acid groups with the -OH group on pretreated $SiO_2$/silicon waveguides surfaces, first demonstrated in this study, can circumvent the aforementioned problem and produce the uniform amine conjugation of crown ethers on waveguide surfaces. We note that the successful Fischer esterification on $SiO_2$/silicon defies the conventional view that the reaction is applicable to organics only[39]. Toward a broader scope, the Fischer esterification of an inorganic material possessing an -OH group implies the agnostic-nature of the process, indicating far-reaching technological implications to replace silylation agents in cases where they are used to couple silica/silicon with organic compounds[37,38]. For instance, different crown ethers[40–50], selective to various ions (i.e., $K^{43}$, $Be^{44}$, $Ra^{46}$, $Cs^{41}$), illustrated at the inset of Fig. 1c, can undergo amine conjugation following Fischer esterification on $SiO_2$/Si, greatly broadening the range of applications (i.e., medical[43], electronics manufacturing[44], nuclear[41,46]) that the developed platform can be extended to. As a corollary of complementary metal-oxide-semiconductor (CMOS) fabrication, SiP has proven to be a disruptive integrated photonic technology that enables high-precision mass manufacturing, without compromises in yield[51–54]. Through the synergistic integration of both technologies, the resulting platform is engineered to overcome several unaddressed issues against lead poisoning in society: 1). The successful amine conjugation of crown ethers via Fischer esterification onto aptly designed SiP circuits will enable *in-situ*, selective, ppb-scale detection of $Pb^{2+}$ ions, improving upon current bulky lab-based systems (ICP-MS/OES)[23,24]. 2.) ICP-MS/OES requires a significant lead time from sample collection to results due to complex lab-based sample processing and analysis[23,24]. The work demonstrates the capacity for rapid, *in operando* determination of $Pb^{2+}$ concentration in analytes, on site. 3). The high-index contrast of silicon against its cladding material ($SiO_2$) enables the design of compact photonic sensors that can be widely applied[55–62]. 4). Through established manufacturing technologies, SiP circuits can be mass produced, at low cost[51–54]. Furthermore, the crown-ether functionalization process is solution-based (with reactants dissolved in green solvents such as water and ethanol), implying that wafer-scale functionalization can be achieved, indicating scalability, with minimal environmental pollution.

## 2. Concept

### 2.1. Photonic Device Design

The $Pb^{2+}$ sensor illustrated in Fig. 1a, is fabricated on the 220 nm silicon-on-insulator platform; a micrograph image is shown in Fig. 1b. Slot waveguides are implemented in the sensing arm ($H_2O$ cladding). As the slot width is comparable to the exponential decay length of the fundamental eigenmode, optical power perpendicular to the high-index contrast interfaces is amplified[60–65]. Essentially, this feature of slot waveguides lends to high surface sensitivity[60–65]. The lightwave propagates to an asymmetric adiabatic tapered splitter[66] (Fig. 2d), where a larger proportion of the optical power is directed to the sensing path. Following, a 250 µm-long strip-to-slot converter is utilized for the transition of the strip to slot optical mode[65] (Fig. 2e). With the exception of the sensing region, which is exposed to the analyte, the entire device is cladded with $SiO_2$. The thickness of the $SiO_2$ cladding is designated to be 2 µm to prevent interaction with the analyte, where > 99 % of the optical power is confined within the boundaries of the cladding. In the reference arm, slot waveguides with identical dimensions are also implemented. The reason for this is to normalize waveguide propagation losses on the sensing and reference arms, and the power ratio of the asymmetrical adiabatic tapered splitters are designed according to water absorption[67] (designed losses) in the sensing region. The lightwave from the sensing and reference arms recombines at the asymmetrical adiabatic tapered splitter, forming the MZI interferometric spectrum. Power ratio of the two splitters is designed to optimize interference fringe visibility[60] (extinction ratio). The operating protocol of the sensor is elucidated in Fig. 1c. First of all, the sensing arm is exposed to deionized (DI) wafer to obtain the reference resonant wavelength ($\lambda_0$); and subsequent wavelength shift will be considered in reference to this wavelength.

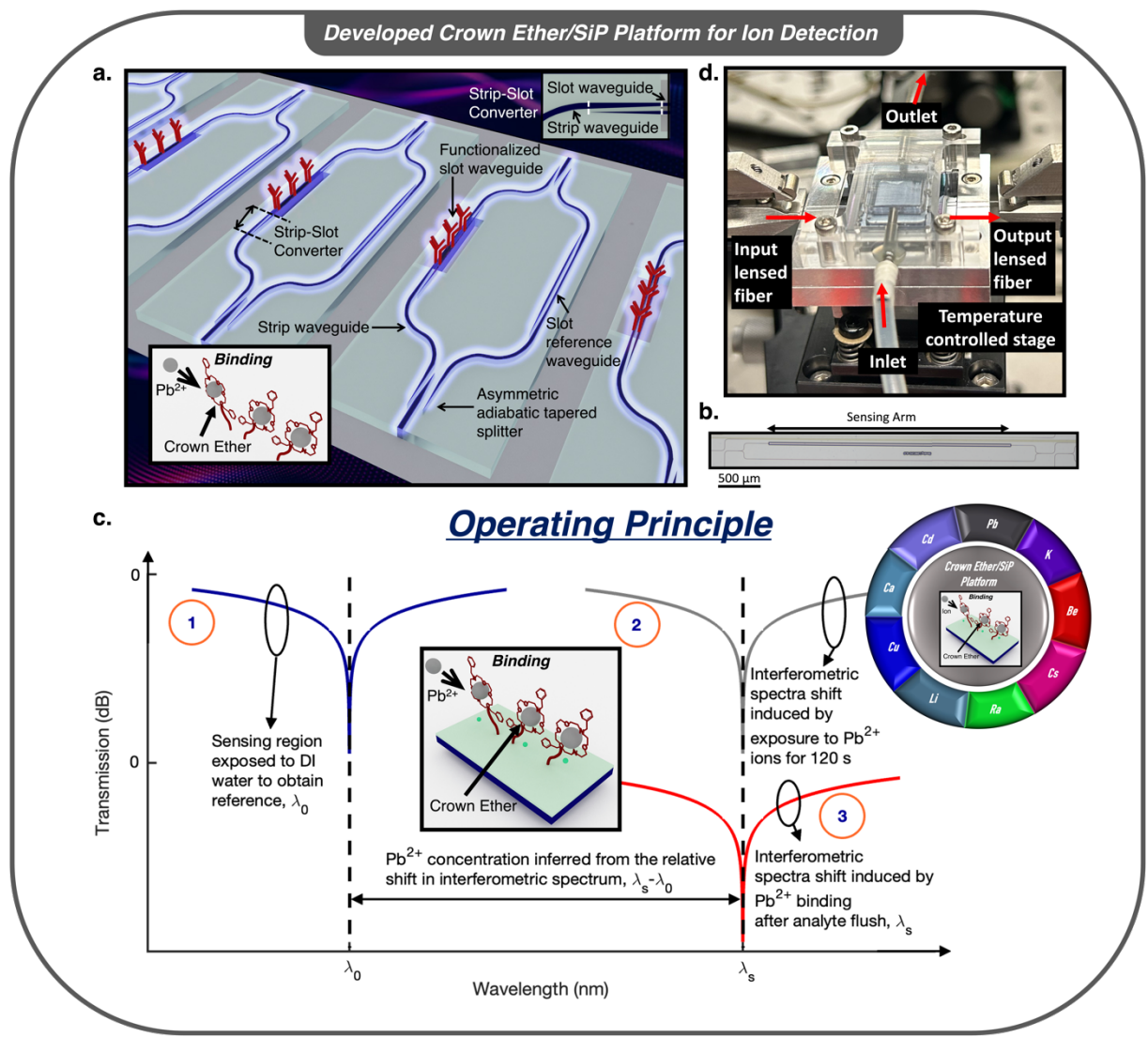

**Fig. 1 Concept of the Crown Ether/SiP platform for ion detection. a,** 3-D illustration of the photonic $Pb^{2+}$ ion sensor based on the crown ether decorated SiP platform. The functionalization performed in the sensing arm is indicated by the features in red, where the inset illustrates the crown ether functionalized to the $Si/SiO_2$ surface by Fischer esterification, and then amine conjugation. **b,** The micrograph image of the $Pb^{2+}$ ion sensor, where the sensing arm and scale bar (500 μm) are indicated. **c,** Elucidated operating principle of the photonic $Pb^{2+}$ ion sensor. The inset shows the exemplary applications that the ion detection platform can be extended to. **d,** The $Pb^{2+}$ photonic sensor assembly, consisting of the photonic chip and a microfluidic chamber.

Following, DI water will be flushed from the microfluidic chamber, and analyte possibly containing $Pb^{2+}$ ions will be added. Should the analyte contain $Pb^{2+}$ ions, a resonant shift will be induced through binding of the ions to the functionalized surface of the sensing region through surface sensing[56,60–62]. However, one should also note that a proportion of the wavelength shift could also be caused by the interaction of the evanescent field with the other particles/ions/molecules in the analyte[56]. Therefore, this necessitates the subsequent flushing of the analyte from the device via the addition of DI water. This results in the retention of $Pb^{2+}$ ions, which are binded to the surface of the sensor through the functionalized layer. Subsequently, the resonant shift ($\lambda_s$) measured can be attributed to the immobilized $Pb^{2+}$ ions. The concentration of the $Pb^{2+}$ ions in the analyte can be deduced from the shift in resonance wavelength from reference ($\lambda_s$-$\lambda_0$), utilizing a calibration curve; the calibration curve indicates wavelength shift as a function of detected concentration as shown later in Fig. 5b. We will henceforth refer to the flushing of analyte and the addition of DI water into the microfluidic chamber after ion interaction as analyte flush.

Fig. 1d shows the photonic chip, with the polydimethylsiloxane (PDMS) microfluidic channel mounted via a stainless-steel fixture. Analyte input and extraction was implemented via the following inlet and outlet tubes. Optical input/output was performed via edge coupling between a lensed fiber with ~3 μm mode field diameter and a silicon coupler that tapers down to 175 nm. The abovementioned assembly (see Supplementary Note 1) was mounted on top of a thermoelectric controller (TEC), maintained at 296 K with thermal drift of lower than 2 mK.

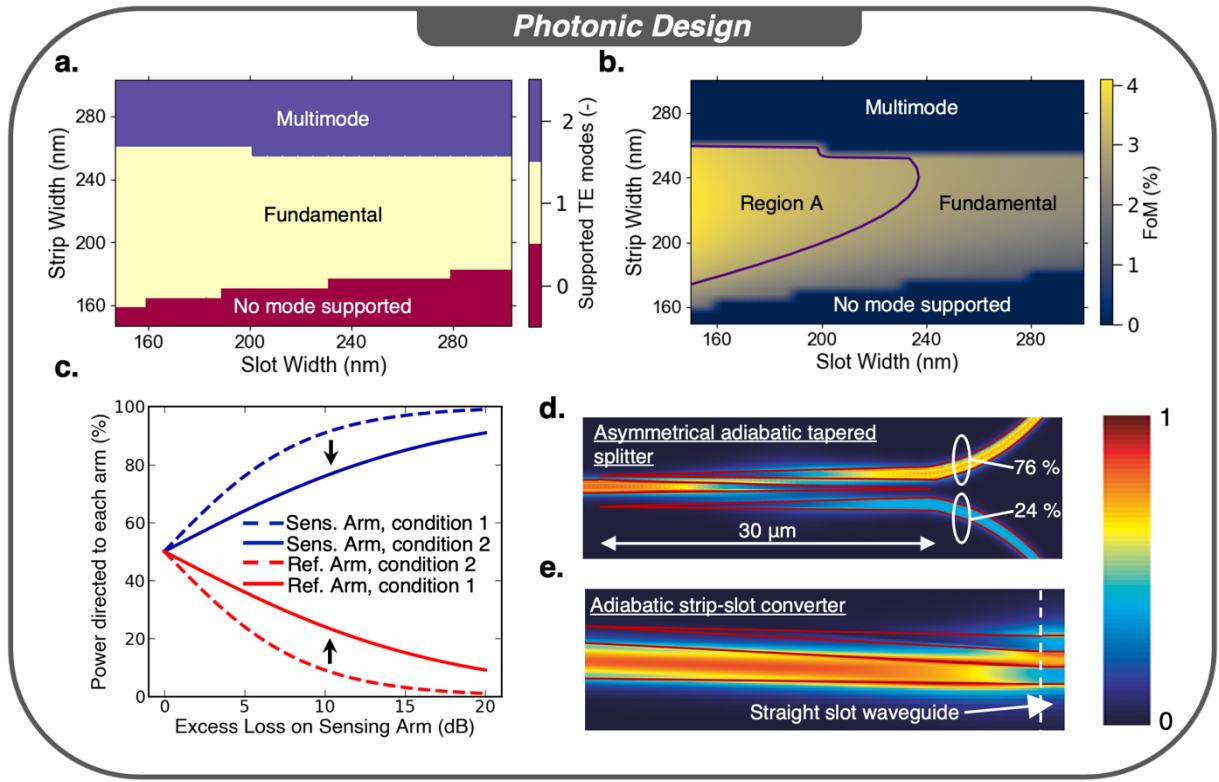

**Fig. 2 Photonic design of the lead ion sensor a,** Simulation of the number of supported TE optical modes in the slot waveguides as a function of strip and slot width. **b,** Sensor surface sensing FoM as a function of strip and slot width. **c,** The comparison of two proposed splitting Mach-Zehnder architectures (see Supplementary Note 3) in terms of the power asymmetry required of the splitter; condition 1 ($S_1 = S_2, S_1' = S_2', S_{1/2}' \neq 0.5$), condition 2 ($S_1' \neq 0.5, S_2' = 0.5$). Top-down electric field distribution of the **d,** asymmetrical adiabatic tapered splitter, and **e,** adiabatic strip-slot converter, where the structure of the components are outlined.

The dimensions of the slot waveguide (Strip and Slot width defined at Supplementary Note 2) were determined via eigenmode calculations in Fig. 2a-b, where $H_2O$ cladding surrounds the structure. First of all, the parameter space corresponding to the number of transverse electric (TE) modes was performed in Fig. 2a. As the top and bottom media surrounding the waveguide is asymmetrical (BOX on the bottom and $H_2O$ as the cladding), there exists a regime where the fundamental mode is not supported (in red); when the mode angle is smaller than the critical angles. Conversely, the multi-mode regime of the slot waveguide structure is indicated in blue, where the second order TE mode will be supported. According to Fig. 2a from left to right, the second order TE mode emerges when the strip and slot widths are ~260 and ~145 nm respectively. In addition, the corresponding strip width that excites the second order TE mode decreases as slot width increases. The parameter space corresponding to single TE mode propagation is highlighted in yellow. For optimization of surface sensitivity, the selection of the optimal strip and slot width, subject to the fundamental TE mode is dependent on the optical mode confinement on the surface of the sensing region. To that effect, a figure-of-merit (FoM) is defined, that takes into consideration, the optical confinement factor within 10 nm about the surface of the slot waveguides which are cladded with 20 nm of $SiO_2$ (see Supplementary Note 2).

Computation of the FoM is performed in the parameter space of Fig 2(a) and the results are presented in Fig. 2b. The corresponding boundary condition for the number of supported TE modes (Fig. 2a) is replicated in Fig. 2b. As a guide to the eye, the highest value of FoM is indicated by region A. However, we have encountered difficulties in the selective removal of $SiO_2$ cladding at the sensing region when the slot gap is smaller than 200 nm. As such, strip and slot width of 240 and 240 nm respectively are implemented to relax process requirements; the selected slot waveguide parameters lie close to the boundary of region A.

$H_2O$ poses significant water absorption at the C-band[67]. Yet, the length of the sensing region increases the surface sensitivity of the sensor; which forces an inherent tradeoff between the fringe visibility (extinction ratio) and sensitivity[56,60–62]. The implementation of asymmetrical splitting in MZIs will serve to overcome the issue. As identical slot waveguide dimensions are implemented on the sensing and reference arms, the primary source of loss difference between the two arms comes from water absorption. It can be concluded that the splitting ratios of the MZIs must be co-designed with the length of the sensing arm, hence designed losses. The quantities are related to one another via the following equation, where the derivation is elaborated (see Supplementary Note 3). Designed losses through water absorption is assumed to be the only source of loss in the sensor.

$$\sqrt{S_1}\sqrt{S_2}e^{-\frac{\alpha L}{2}} = \sqrt{S_1'}\sqrt{S_2'} \quad (3)$$

$S_1, S_1'$ and $S_2, S_2'$ refers to the splitting ratios of the input and output splitter respectively. Assuming the splitters are lossless, energy conservation dictates that $S_{1/2} = 1 - S_{1/2}'$. $\alpha$ is the loss coefficient due to water absorption, and $L$ refers to the length of the sensing arm. We propose a condition such that $S_1 = S_2, S_1' = S_2'$ and $S_{1/2}' \neq 0.5$ (condition 1). In Fig. 2c, we plot the splitting ratios to the sensing and reference arm corresponding to maximum visibility. A comparison to an alternate condition where arbitrary splitting is at the input splitter, and 3-dB splitting is at the output splitter (condition 2, $S_1' \neq 0.5, S_2' = 0.5$) is also indicated in Fig. 2c; see Supplementary Note 3. In comparison, condition 1 reduces the asymmetry that is required of the splitters, alleviating fabrication requirements. As a compromise between the sensor surface sensitivity and the optical measurement setup power budget, our demonstration selected splitting ratios corresponding to sensor arm designed loss of 10 dB: $S_1 = 0.76, S_2 = 0.24$. In regard to the selected slot waveguide dimensions in a water cladding, the waveguide propagation loss due to $H_2O$ absorption is estimated to be 35 dB/cm at $\lambda$ = 1.55 μm. This gives rise to a sensing arm length of $\frac{Designed\ Loss\ (dB)}{Propagation\ Loss\ (dB/cm)} = \frac{10}{35} = 2857.14\ \mu m$.

Asymmetrical adiabatic tapered splitter, which has been developed in our previous work[66], are implemented for arbitrary power splitting. These power splitter offers the advantage of broadband operation and low loss. In Fig. 2d, we show the top-down electrical field distribution of a 30 μm-long 76/24 % power splitter. The length of the strip-to-slot converter is 250 μm, where low-loss adiabatic conversion from strip to slot mode is facilitated[65]. Similarly, the top-down electric field distribution of the lightwave as it propagates along the converter is indicated in Fig. 2e.

## 2.2 Functionalization and Characterization of 7,16-Dibenzyl-1,4,10,13-tetraoxa-7,16-diazacyclooctadecane (DBTDA) functionalized chip via Fischer Esterification and Amine Conjugation

The functionalization protocol can be divided into four steps (Fig. 3a) with purification implemented after every step. Firstly, a layer of $SiO_2$ (20 nm) is deposited onto silicon via Atomic Layer Deposition (ALD) in the pretreatment step. Secondly, the Fischer esterification method[39] is used to couple the silanol group on the $SiO_2$ surface of the SiP slot waveguides (in the sensing region) with polyacrylic acid, in the presence of a tin catalyst[68]; specifically tin (II) chloride dissolved in DI water. The photonic chips, immersed in the reagent, was maintained at 318 K and agitated, while the reaction proceeds for 60 minutes. The chips are then cleaned with DI water. Following, the photonic chips are functionalized with DBTA, forming amide bonds via conjugation, in the presence of 1-Ethyl-3-(3-dimethylaminopropyl)-carbodiimide), EDC, and N-Hydroxysuccinimide NHS dissolved in ethanol[24,69]. In the final step, the functionalized photonic chips are rinsed in dilute nitric acid to remove as much of tin catalyst adsorbed on the surface as possible. This was done by placing the functionalized photonic chips in a solution of 0.1 M nitric acid, agitated with a magnetic stir bar for 60 minutes.

X-ray Photoelectron Spectroscopy (XPS) analysis was used to characterize the functionalized photonic chips before functionalization as well as, before and after interaction with different ions to analyze the resulting elemental constitution. A comparison of the N 1s region[70] of the XPS spectrums before and after functionalization (Fig. 3b) shows that the presence of nitrogen is only detected after the functionalization process. While the comparison of the C 1s region[70] of the XPS spectrums before and after functionalization (Fig. 3c) shows a larger peak area and chemical shift of C1s carbon towards a higher binding energies (~286.5 eV[70], ~288.8 eV[70]). The binding energy level at 286.5 eV indicates higher carbon concentration and significant carbon binding with electronegative species such as oxygen (C-O) and nitrogen (C-N)[70], and the peak at ~288.8 eV corresponds to C=O[70]. Furthermore, the comparison of the O 1s region[70] of the XPS spectra before and after functionalization (Fig. 3 d) shows a decrease in oxygen signal and chemical shift of O 1s towards a lower binding energy.

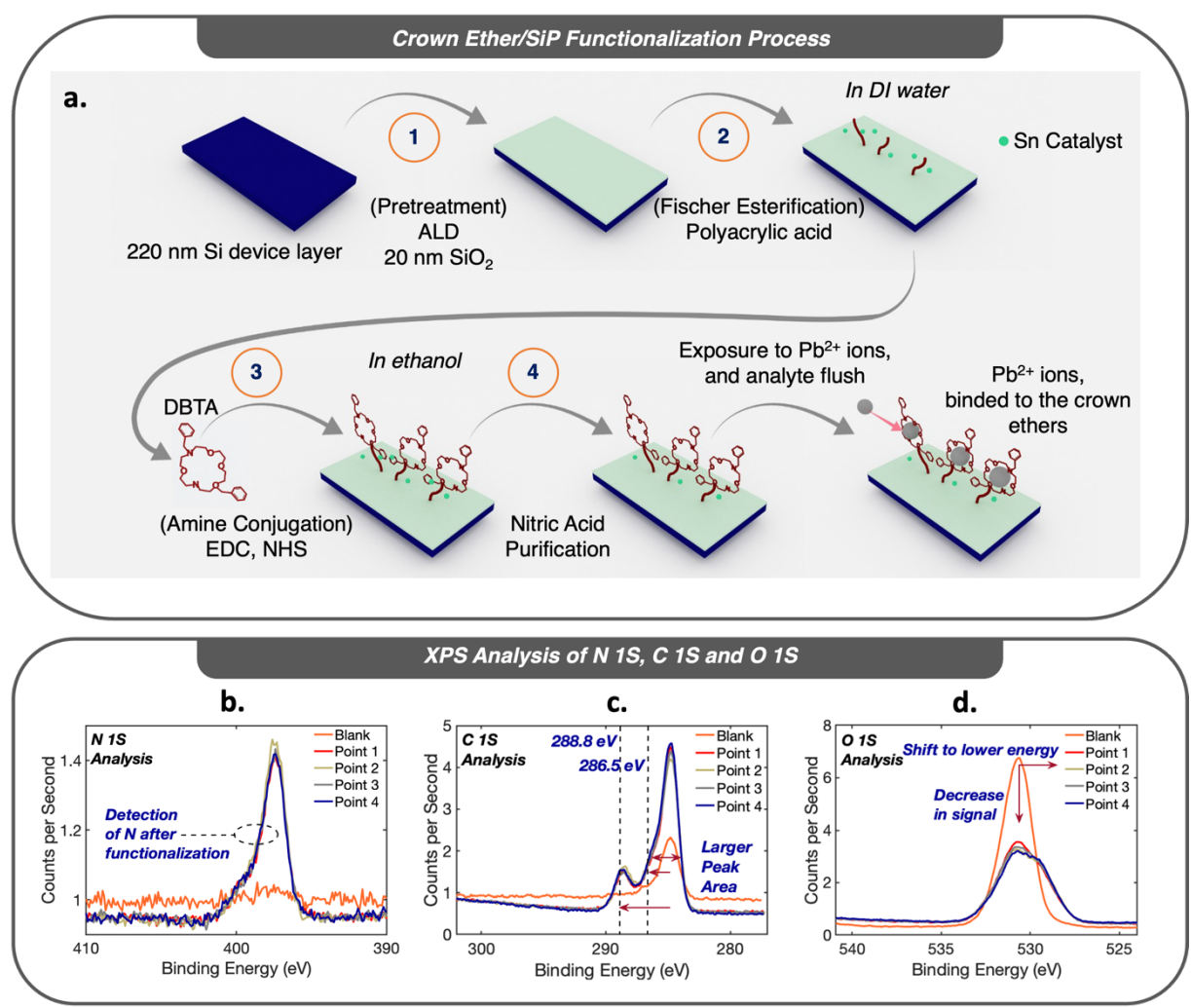

**Fig. 3 The development of the Crown Ether/SiP functionalization process a,** The developed crown ether/SiP functionalization process, described in 4 steps. XPS narrow spectra analysis of the **b,** N 1S, **c,** C 1S, **d,** O 1S regions of the photonic chips, before and after functionalization.

This observation can be attributed to the functional layer consisting of more carbon compared to oxygen[70], and the Fischer esterification of silanol with the carboxylic acid group forms O-C bonds which have a lower binding energy compared to O-Si bonds[70]. The above points to significant evidence that successful Fischer esterification have been achieved following the elucidated protocol (Fig. 3a). Furthermore, to assess the uniformity of functionalization, in Fig. 3b-d, the N 1s, C 1s and O 1s regions respectively of the XPS spectra were taken at four different points (Point 1, 2, 3, 4) on the photonic chip spaced more than 1 cm apart on the photonic chip, showing high consistency. This implies that a functional layer with good uniformity have been realized. In addition, Energy Dispersive X-ray (EDX) analysis pertaining to elemental analysis of the functional layer was performed and included in Supplementary Note 4. The results provide compelling evidence for the successful uniform functionalization of the DBTA crown ether, which contains carbon bonded to nitrogen and oxygen, via Fischer esterification and amine conjugation.

Na, K, Mg, Li, Zn, Ca, Fe, Cu, Al, Sn, Cd, and Pb were chosen as highly-relevant analytes to quantify the selectivity of the photonic-based ion detection platform. The selected ions demonstrates a variety of ionic sizes and charge. Na, K, Ca, Mg, Zn and Cu are commonly found in bottled water sources[71], while Fe, Li, and Al could be present in groundwater sources[72]. Sn is used as a catalyst in the Fischer esterification process[68], and Cd and Pb[73] are toxic heavy metals that should be prohibited in drinking water. Each functionalized photonic chip interacts with 100 ppb of the abovementioned analyte independently for 120 s in a microfluidic chamber. After which, the analyte is flushed with DI water and dried with $N_2$ gas blow. XPS is utilized to identify the elemental constitution on the surface of the photonic chips before and after ion interaction via the respective elemental binding energies of each element. Normalization was carried out where the narrow scan XPS spectra prior to ion interaction was subtracted from that after ion interaction. The normalized narrow scan XPS spectra of Na, K, Mg, Li, Zn, Ca, Fe, Cu, Al, Sn, and Cd are displayed in Fig. 4a–k respectively, indicating the absence of binding on the functionalized photonic sensor. For the abovementioned ions, we note that only $Sn^{2+}$, which is used as the catalyst during Fischer esterification, have been identified prior to ion interaction (see Supplementary Note 5). Conventionally, Fischer esterification is favored when $H_2O$ is removed as the reaction proceeds (dehydrative esterification). However, the developed esterification process in this work utilizes $H_2O$ as a green solvent, which will decrease the catalytic activity of Brønsted acid catalysts (i.e., $H_2SO_4$)[74]. In the process (Fig. 3a), the $H_2O$-tolerant Lewis acid catalyst $SnCl_2$, is used[68,75] where Sn is embedded into the $SiO_2$, functioning as a heterogeneous catalyst in the process. It is known that heterogeneous catalysts show improved catalytic activity[76] that favors esterification even in the presence of $H_2O$[77]. This is verified in Fig. 3b-d and Supplementary Fig. 4 (see Supplementary Note 4). In Fig. 4i, unmistakable binding of $Pb^{2+}$ ions is demonstrated, indicating the presence of $Pb^{2+}$ binding events on the functional layer, via identification of the Pb $4f_{5/2}$ and Pb $4f_{7/2}$ elemental binding energies[70]. Furthermore, in Supplementary Note 4, EDX analysis is performed, where the absence and presence of $Pb^{2+}$ can be clearly seen before and after interaction respectively. From the above, it can be anticipated that the photonic sensor will be selective only towards $Pb^{2+}$, where the ion will bind to the functionalized surface, and be present after analyte flushing. Subsequently, the concentration of exposed $Pb^{2+}$ can be inferred from photonic surface sensing via the shift in the interferometric spectrum.

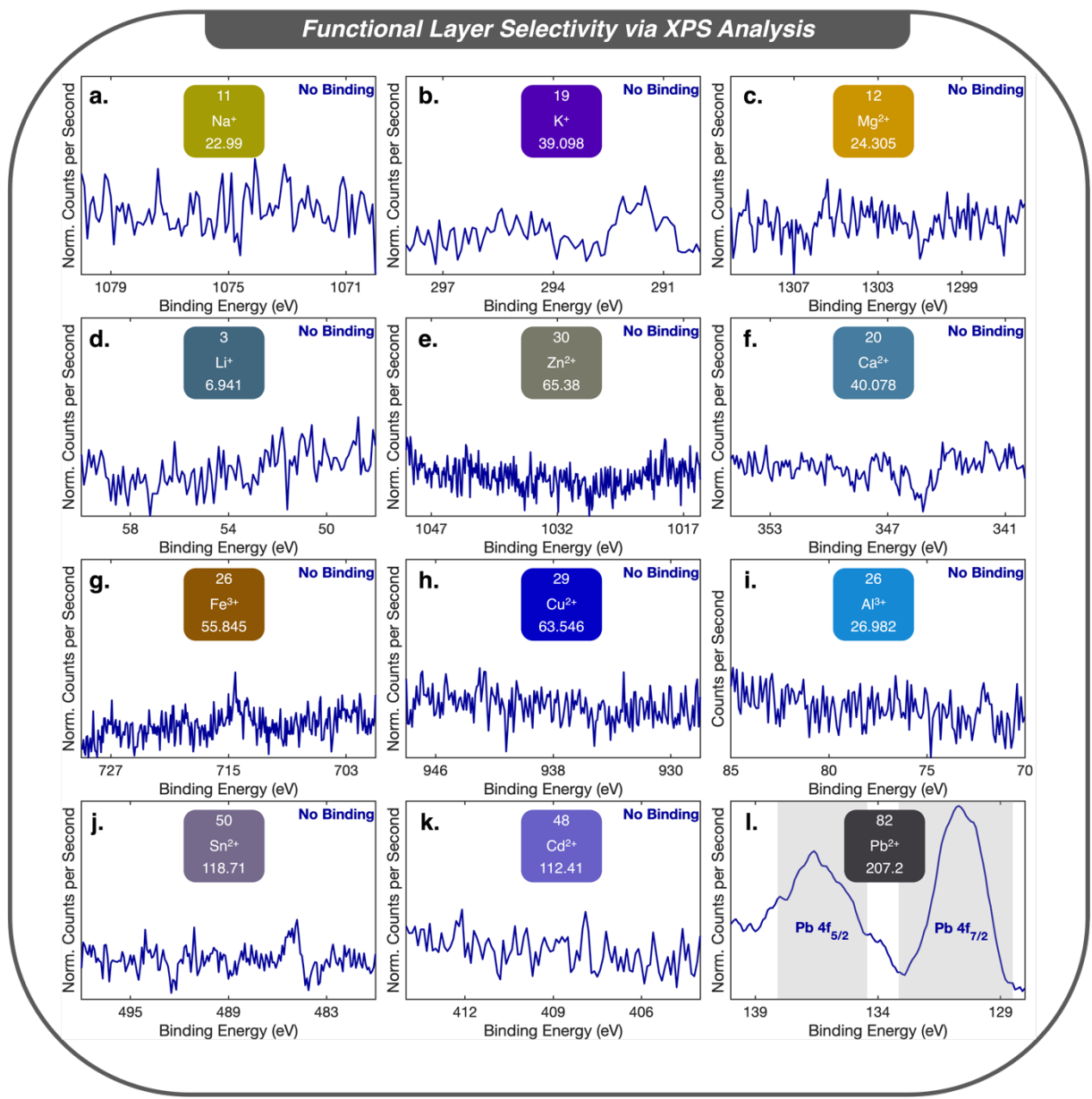

**Fig. 4 Analysis of functional layer selectivity via XPS.** Normalized narrow scan XPS spectra at photonic chip surface by subtracting the spectra prior to ion interaction from that of after ion interaction **a,** Na$^+$, **b,** K$^+$, **c,** Mg$^+$, **d,** Li$^+$, **e,** Zn$^{2+}$, **f,** Ca$^{2+}$, **g,** Fe$^{2+}$, **h,** Cu$^{2+}$, **i,** Al$^{3+}$, **j,** Sn$^{2+}$, **k,** Cd$^{2+}$, and **l,** Pb$^{2+}$ ions respectively.

## 3. Pb$^{2+}$ Photonic Sensor Characterization

In Fig. 5a, we show the measured fundamental TE mode transmission of the functionalized photonic sensor, exposed to DI water; see Methods for maintenance of fundamental TE. Details pertaining to SiP chip fabrication is elaborated in Supplementary Note 7. As predicted, good interference fringe visibility is obtained, at λ = 1531.9 nm, where extinction ratio exceeds 20 dB. In contrast, when the sensing region is exposed to air, poor visibility is observed (see Supplementary Note 6). It is observed that sensor visibility is reduced at λ = 1563.7 nm, attributed by the lower water absorption at λ = 1563.7 nm. The experimental demonstration of the sensor visibility is limited by fabrication bias from design parameters. This would induce changes in designed splitting ratios ($S_1, S_1', S_2, S_2'$), as well as loss difference between the reference and sensing arms. The measured free-spectral range (FSR) is 31.7 nm.

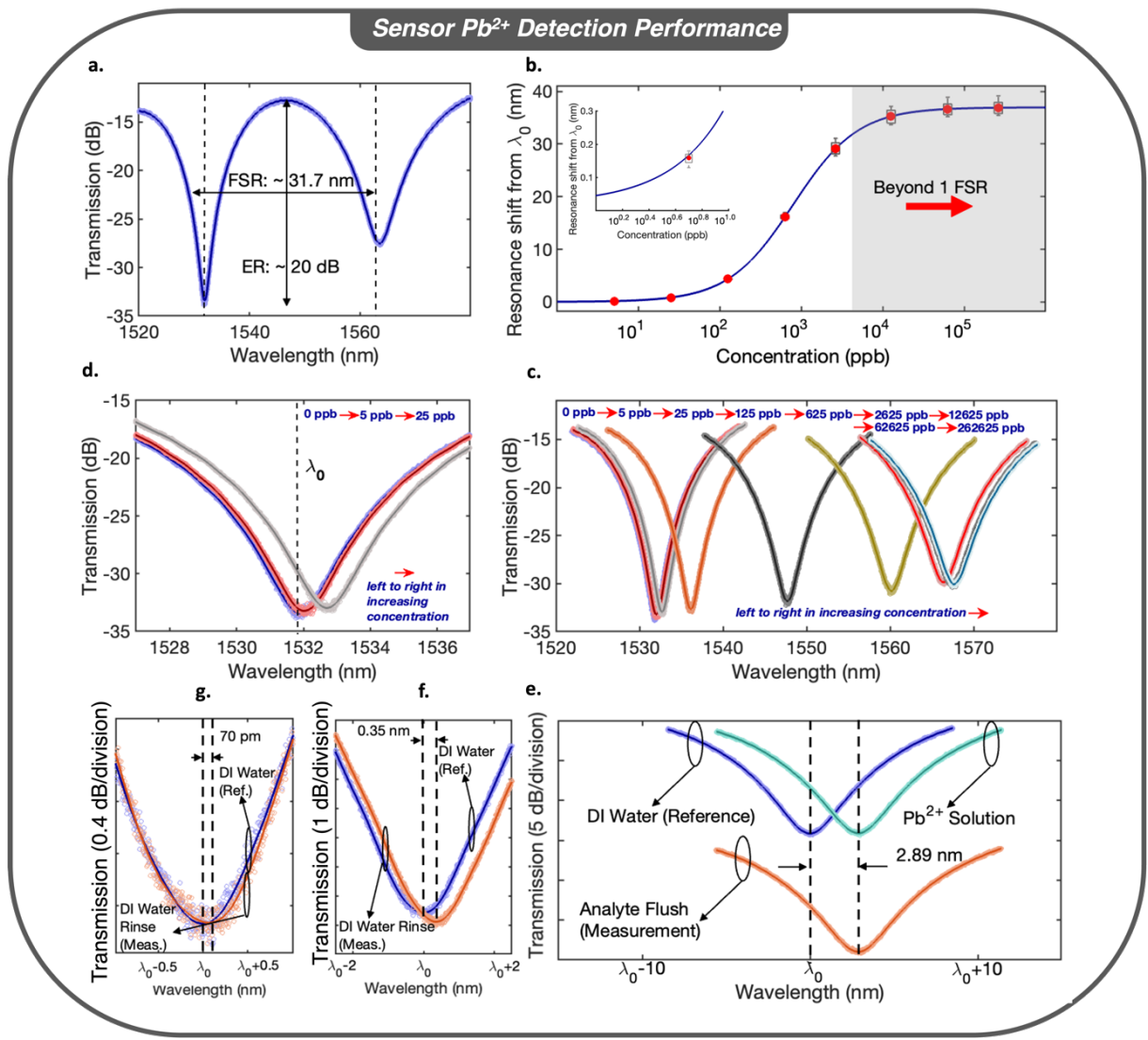

**Fig. 5. Experimental characterization of Pb²⁺ sensor performance. a,** Wavelength spectrum of the sensor, when DI water is applied into the sensor assembly (as shown in Fig. 1c). **b,** Calibration curve of the sensor when exposed to $Pb^{2+}$ concentration of 0, 5, 25, 125, 625, 2625, 12625, 62625, 262625 ppb, via a cumulative testing approach (see Experimental Section). **c,** A set of fringe minima corresponding to the tested concentration in the calibration curve (Fig. 5b). **d,** Fig. 5c, zoomed-in at concentrations of 0, 5, 25 ppb. Validation of the calibration curve (Fig. 5b) at concentrations of **e,** 80, **f.** 10 and **g,** 1 ppb.

The calibration curve of the $Pb^{2+}$ sensor, indicating resonant wavelength shift as a function of cumulative $Pb^{2+}$ concentration within the analyte (5, 25, 625, 2625, 62625, 262625 ppb) is indicated in Fig. 5b through a cumulative testing process (see Methods); the inset shows the resonant wavelength shift within a range of 1 to 10 ppb. Interpolation was carried out to understand the form of the calibration curve. The sigmoidal curve is characteristic of absorption isotherms[78,79]. As indicated by the shaded section in Fig. 5b, the wavelength shift exceeds a single FSR when the cumulative concentration of $Pb^{2+}$ is higher than ~60000 ppb. In Fig. 5c, we show a set of MZI spectra around the minima transmission points for the fringes corresponding to each of the abovementioned concentrations. As a large cumulative concentration range is presented in Fig. 5c, the spectra when cumulative $Pb^{2+}$ concentration are 0, 5 and 25 ppb are shown (Fig. 5d). A saturation of $\lambda_s - \lambda_0$ against cumulative concentration is observed in Fig. 5b. This is ascribed to the saturation of the binding sites within the functional layer[78,79].

To affirm the reproducibility of the calibration curve in Fig. 5b, three photonic sensors were tested independently, at concentrations of 80, 10 and 1 ppb. In Fig. 5e, the transmission spectra when the sensing region is exposed to DI water, DI water containing 80 ppb $Pb^{2+}$, as well as after analyte flush of $Pb^{2+}$ are shown. Based on the positions of $\lambda_s$ in comparison to $\lambda_0$, a $\lambda_s - \lambda_0$ of 2.89 nm is obtained. This shift was observed after analyte flushing, consistent with the binding mechanism on the surface of the sensor. By comparing this value to Fig. 5b, the inferred concentration from the calibration curve is 81.65 ppb, which is close to the actual value of 80 ppb. Similarly, Fig. 5f and 5g yield concentrations of 10.56 and 1.7 ppb as determined from the calibration curve, versus ground truths of 10 and 1 ppb respectively. The accuracy of the photonic measurement setup in determining the resonant wavelength of the sensor interferometric spectrum is ~20 pm, obtained from multiple spectral measurements of a sensor, at the same condition (10 ppb $Pb^{2+}$ followed by analyte flush). This indicates that the sensor is capable of detecting $Pb^{2+}$ at concentrations much lower than the EPA standard.

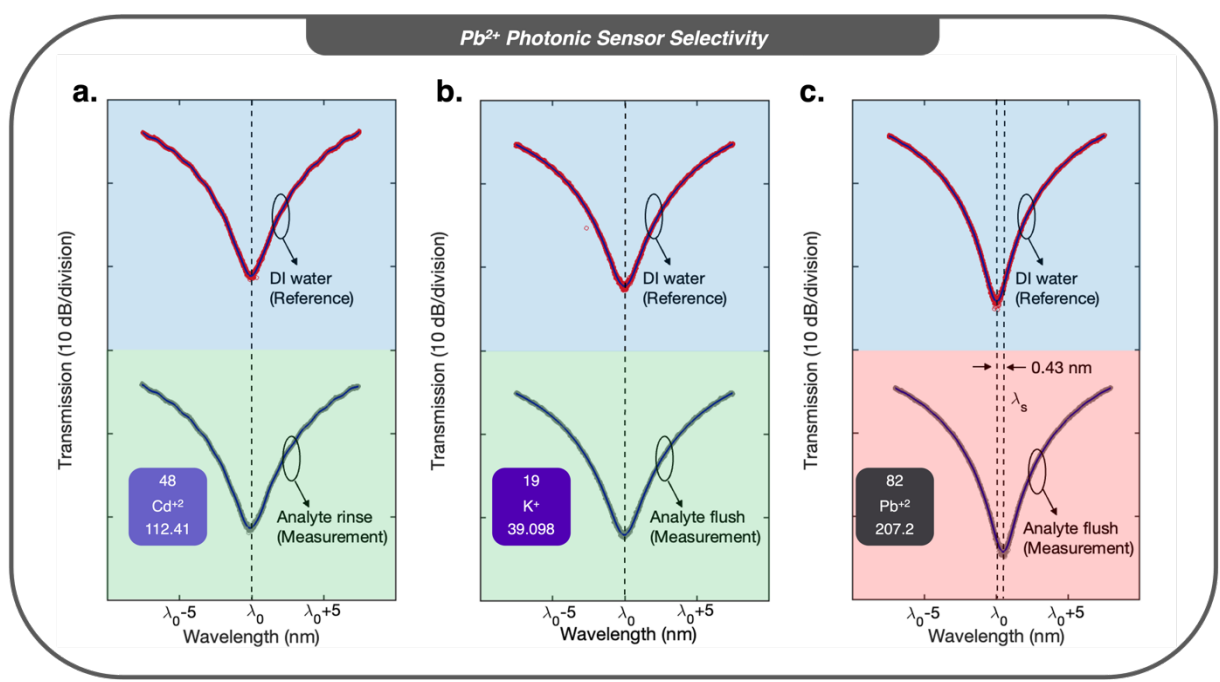

**Fig. 6. Selectivity performance of the Pb$^{+2}$ ion photonic sensor against a,** Cd$^{2+}$, and **b,** K$^+$ at 15 ppb where no shifts in the interferometric spectra indicative of ion binding is observed. Similar to Fig. 5 e-g, the detection performance of the Pb$^{2+}$ photonic ion sensor is tested at 15 ppb, where significant ion binding, resulting in interferometric shifts that corresponds closely to Fig. 5b is observed.

In Fig. 4 (XPS analysis), the functional layer is found to be selective to Pb$^{+2}$ ions against the other tested ions, which implies the selectivity of the photonic sensor. To further verify sensor selectivity performance at the Pb$^{+2}$ safety threshold (15 ppb[19]) Na$^+$, K$^+$, Mg$^{2+}$, Ca$^{2+}$, Fe$^{3+}$, Cu$^{2+}$, Sn$^{2+}$, Cd$^{2+}$ are all tested at 15 ppb; Cd$^{2+}$ and K$^+$ are shown in Fig. 6a-b respectively. As expected from Fig. 4, no shift in the interferometric spectrum indicative of ion binding is observed. The reference interferometric spectrum (in DI water), and the resulting spectrum after exposure and analyte flush for Cd$^{+2}$ and K$^+$ are presented in Fig. 6a-b respectively and data for other ions are presented in Supplementary Note 8. The data indicates no shift in the interferometric spectrum indicative of ion binding. Similar to Fig. 5e-g, the photonic sensor was tested at 15 ppb of Pb$^{2+}$, where a shift of 0.43 nm corresponding to 13.1 ppb is obtained, in reference to the calibration curve in Fig. 5b. The results in Fig. 6 and Supplementary Note 8 underpins the ability of the sensor to effectively detect Pb$^{+2}$ ions in the presence of the other ions. A more comprehensive selectivity study is required, and is currently being conducted.

## 4. Conclusion

In this work, for the first time, crown ether functionalization via Fischer esterification and subsequent amine conjugation is integated with highly-scalable, and low cost inorganic SiP. This realizes a photonic platform that enables the selective binding of Pb$^{2+}$ ions, and subsequent detection down to the ppb-scale. The novel reaction pathway proposed and demonstrated, driven via Fischer esterification defies prior expectations that the process is restricted to organics[39]. This enables the engineering of the platform to selectively detect a plethora of ions via subsequent amine conjugation of various crown ethers[40–50]. Furthermore, the functionalization process, by virtue of being solution-based, can be implemented at the wafer-scale. The reactants are dissolved in green solvents which results in minimal environmental impact. The sensor presented in this work indicates the ability to detect Pb$^{2+}$ concentrations *in-situ*, through a wide dynamic range (1 – 262000 ppb) while being highly-selective against other commonly-found, relevant ions. This work represents an encouraging step towards the ubiquitous implementation of photonic-based sensors that protects against widespread Pb$^{2+}$ poisoning. We envisage that this platform can be extended to multiplex ion detection in multiple application spheres.

**Methods**

***Microfluidic chamber fabrication***: A custom made acrylic top enclosure, polydimethylsiloxane (PDMS) gasket and bottom mount make up the flow channel assembly (see Supplementary Note 1). This allows the sample solution and DI water to flow across the sensor on the photonic chips, and doubles as a containment to allow a fixed volume of sample solution to stay atop the sensor for 120 s during ion interaction. The custom made PDMS gasket is fabricated by curing a PDMS and photo-initiator mixture (Shin-Etsu KER-4690) in polytetrafluoroethylene (PTFE) mold under 405 nm UV lamp for 10 minutes.

***Analyte preparation***: The analyte solution preparation is carried out by diluting 1000 ppm ICP standard solutions of the selected ions (Merck) with DI water to the concentration required; for low concentrations (lower than 10 ppm), multiple rounds of dilution were performed. ICP-MS is used to verify the concentrations.

***Fundamental TE maintenance***: The $Pb^{2+}$ photonic sensor is designed for fundamental TE operation. Fundamental TE operation is crucial for the maintenance of interference fringes corresponding to the mode as well as fringe visibility. In order to ensure that the device is operating with only the fundamental TE mode, we utilized a chain of cascaded Multi-Mode Interferometer (MMI) structures that is optimized for the desired polarization (10 × MMI). The polarization dependent loss the TM mode experience over TE is 2 dB per MMI. Cascading 10 MMIs yields a TM against TE polarization extinction ratio of 20 dB. By optimizing the input polarization corresponding to the maximum optical power at the output, we will be able to ensure that the device operates with only the fundamental TE mode.

***Measurement of the calibration curve via a cumulative approach***: The following is implemented for the measurement of the calibration curve. DI water was first added to obtain the reference wavelength ($\lambda_0$), and then flushed from the microfluidic chamber (Step 1). Next, DI water containing $Pb^{2+}$ was added and held for 120 s to facilitate binding of $Pb^{2+}$ to the functional surface at the sensing region (Step 2). The analyte was then removed again and DI water is added, where the resonant wavelength is measured to remove the unbound species (Step 3). The optical transmission was subsequently measured. The shift in wavelength is determined by $\lambda_s - \lambda_0$. For the 6 concentrations that were measured (5, 25, 625, 2625, 62625, 262625 ppb), an additive approach was used. In Step 2, DI water containing 5, 20, 600, 2000, 60000, 200000 ppb of $Pb^{2+}$ is added sequentially as the exposed concentration is increased. The measurement was repeated six times for each set of concentrations involving each photonic sensors, as indicated by the associated error bars in Fig. 5(b).

**Data Availability**

The datasets generated and analyzed in the current study are available from the corresponding authors upon reasonable request.

**Acknowledgements**

Jia Xu Brian Sia would like to acknowledge Ministry of Education/NTU College of Engineering International Postdoctoral Fellowship. This work was performed in part through the use of the facilities of MIT.nano, Harvard University Center for Nanoscale Systems (CNS), NTU's Centre for Micro- & Nano-Electronics (CMNE) and Nanyang Nanofabrication Centre (N2FC). The authors also wish to acknowledge this collaborative work with Fingate Technologies and Vulcan Photonics.

**Figure Legends**

**Fig. 1 Concept of the Crown Ether/SiP platform for ion detection. a,** 3-D illustration of the photonic $Pb^{2+}$ ion sensor based on the crown ether decorated SiP platform. The functionalization performed in the sensing arm is indicated by the features in red, where the inset illustrates the crown ether functionalized to the $Si/SiO_2$ surface by Fischer esterification, and then amine conjugation. **b,** The micrograph image of the $Pb^{2+}$ ion sensor, where the sensing arm and scale bar (500 μm) are indicated. **c,** Elucidated operating principle of the photonic $Pb^{2+}$ ion sensor. The inset shows the exemplary applications that the ion detection platform can be extended to. **d,** The $Pb^{2+}$ photonic sensor assembly, consisting of the photonic chip and a microfluidic chamber.

**Fig. 2 Photonic design of the lead ion sensor a,** Simulation of the number of supported TE optical modes in the slot waveguides as a function of strip and slot width. **b,** Sensor surface sensing FoM as a function of strip and slot width. **c,** The comparison of two proposed splitting Mach-Zehnder architectures (see Supplementary Note 3) in terms of the power asymmetry required of the splitter; condition 1 ($S_1 = S_2, S_1' = S_2', S_{1/2}' \neq 0.5$), condition 2

($S_1' \neq 0.5, S_2' = 0.5$). Top-down electric field distribution of the **d,** asymmetrical adiabatic tapered splitter, and **e,** adiabatic strip-slot converter, where the structure of the components are outlined.

**Fig. 3 The development of the Crown Ether/SiP functionalization process a,** The developed crown ether/SiP functionalization process, described in 4 steps. XPS narrow spectra analysis of the **b,** N 1S, **c,** C 1S, **d,** O 1S regions of the photonic chips, before and after functionalization.

**Fig. 4 Analysis of functional layer selectivity via XPS.** Normalized narrow scan XPS spectra at photonic chip surface by subtracting the spectra prior to ion interaction from that of after ion interaction **a,** $Na^+$, **b,** $K^+$, **c,** $Mg^+$, **d,** $Li^+$, **e,** $Zn^{2+}$, **f,** $Ca^{2+}$, **g,** $Fe^{2+}$, **h,** $Cu^{2+}$, **i,** $Al^{3+}$, **j,** $Sn^{2+}$, **k,** $Cd^{2+}$, and **l,** $Pb^{2+}$ ions respectively.

**Fig. 5. Experimental characterization of $Pb^{2+}$ sensor performance. a,** Wavelength spectrum of the sensor, when DI water is applied into the sensor assembly (as shown in Fig. 1c). **b,** Calibration curve of the sensor when exposed to $Pb^{2+}$ concentration of 0, 5, 25, 125, 625, 2625, 12625, 62625, 262625 ppb, via a cumulative testing approach (see Experimental Section). **c,** A set of fringe minima corresponding to the tested concentration in the calibration curve (Fig. 5b). **d,** Fig. 5c, zoomed-in at concentrations of 0, 5, 25 ppb. Validation of the calibration curve (Fig. 5b) at concentrations of **e,** 80, **f.** 10 and **g,** 1 ppb.

**Fig. 6. Selectivity performance of the $Pb^{+2}$ ion photonic sensor against a,** $Cd^{2+}$, and **b,** $K^+$ at 15 ppb where no shifts in the interferometric spectra indicative of ion binding is observed. Similar to Fig. 5 e-g, the detection performance of the $Pb^{2+}$ photonic ion sensor is tested at 15 ppb, where significant ion binding, resulting in interferometric shifts that corresponds closely to Fig. 5b is observed.

**Supplementary Fig. 1.** Exploded view representation of the $Pb^{2+}$ ion sensor assembly, consisting of the bottom mount, photonic chip, PDMS gasket, acrylic top enclosure, and the screws, which are used to secure the assembly, and prevent microfluidic chamber leakage. The assembly is mounted on top of a TEC controller for thermal stabilization.

**Supplementary Fig. 2. Definition of FoM for the optimization of waveguide surface sensitivity. The height of the waveguide layer is 220 nm with a 20 nm $SiO_2$ layer surrounding the waveguide, deposited via ALD. a,** Cross-sectional schematic of the slot waveguide with definition of slot and strip width illustrated; the area of confinement is illustrated. **b,** An instance of electric field distribution pertaining to a slot waveguide with slot and strip width of 240 nm. **c,** Cross-sectional schematic of the strip waveguide(TE/TM) with definition of strip width illustrated; the area of confinement is illustrated. **d,** Sensor surface sensing FoM for strip waveguides (TE and TM) as a function of strip width. Instances of electric field distribution pertaining to **e,** TE strip waveguide with width of 270 nm, and **f,** TM strip waveguide with width of 425 nm.

**Supplementary Fig 3. Architectures of two MZI-based sensor designs with a,** Condition 1, $S_1 = S_2, S_1' = S_2', S_{1/2} = 1 - S_{1/2}'$, where $S_{1/2}' \neq 0.5$, and **b,** Condition 2, $S_{1/2} = 1 - S_{1/2}'$, where $S_1' \neq 0.5$, and $S_2' = 0.5$. According to Fig. 2c of the main text, it can be seen that the architecture in Supplementary Fig. 2a, condition 1 imposes lower requirements on splitter asymmetry as compared to Supplementary Fig. 2b, condition 2.

**Supplementary Fig. 4 EDX analysis of the photonic chips pertaining to Si, N, C, O, and Pb elemental composition a,** prior functionalization, **b,** after functionalization, and **c,** after $Pb^{2+}$ exposure and rinse with DI water and dry ($N_2$).

**Supplementary Fig. 5** XPS spectrum measured after dilute nitric acid purification and prior to $Sn^{2+}$ exposure, and after Sn exposure with DI water rinse and dry ($N_2$).

**Supplementary Fig. 6. Measured optical spectrum of the photonic sensor when the sensing region of the photonic sensor is exposed to air.** Poor visibility, as implied from the interferometric spectrum ER is observed, which is resultant when the designed losses are non-optimal to the asymmetrical splitting ratios.

**Supplementary Fig. 7** The SEM image (false colour) of fabricated slot waveguide, with strip and slot widths of 240 nm.

**Supplementary Fig. 8 Selectivity performance of the $Pb^{+2}$ ion sensor against a,** $Na^+$, **b,** $Mg^+$, **c,** $Cu^+$, **d,** $Sn^{+2}$, **e,** $Ca^{+2}$, **f,** $Fe^{+2}$ at 15 ppb where no shifts in the interferometric spectra indicative of ion binding is observed across all the tested ions.

# Crown ether decorated silicon photonics for safeguarding against lead poisoning

**Luigi Ranno[1,†], Yong Zen Tan[2,†], Chi Siang Ong[2], Xin Guo[3], Khong Nee Koo[4], Xiang Li[3], Wanjun Wang[3], Samuel Serna[1], Chongyang Liu[5], Rusli[3], Callum G. Littlejohns[6], Graham T. Reed[6], Juejun Hu[1], Hong Wang[3] and Jia Xu Brian Sia[1,3,\*]**

[1]Department of Materials Science & Engineering, Massachusetts Institute of Technology, Cambridge, M.A., USA

[2]Fingate Technologies, Singapore

[3]School of Electrical and Electronic Engineering, Nanyang Technological University, 50 Nanyang Avenue, 639798, Singapore

[4]Vulcan Photonics, Kuala Lumpur, Malaysia

[5]Temasek Laboratories, Nanyang Technological University, 50 Nanyang Avenue, 637553, Singapore

[6]Optoelectronics Research Centre, University of Southampton, Southampton SO17 1BJ, UK

[†]These authors contributed equally

*Corresponding author: jxbsia@mit.edu, jiaxubrian.sia@ntu.edu.sg

**Supplementary Note 1: Exploded-view schematic representation of the Pb$^{2+}$ ion sensor assembly**

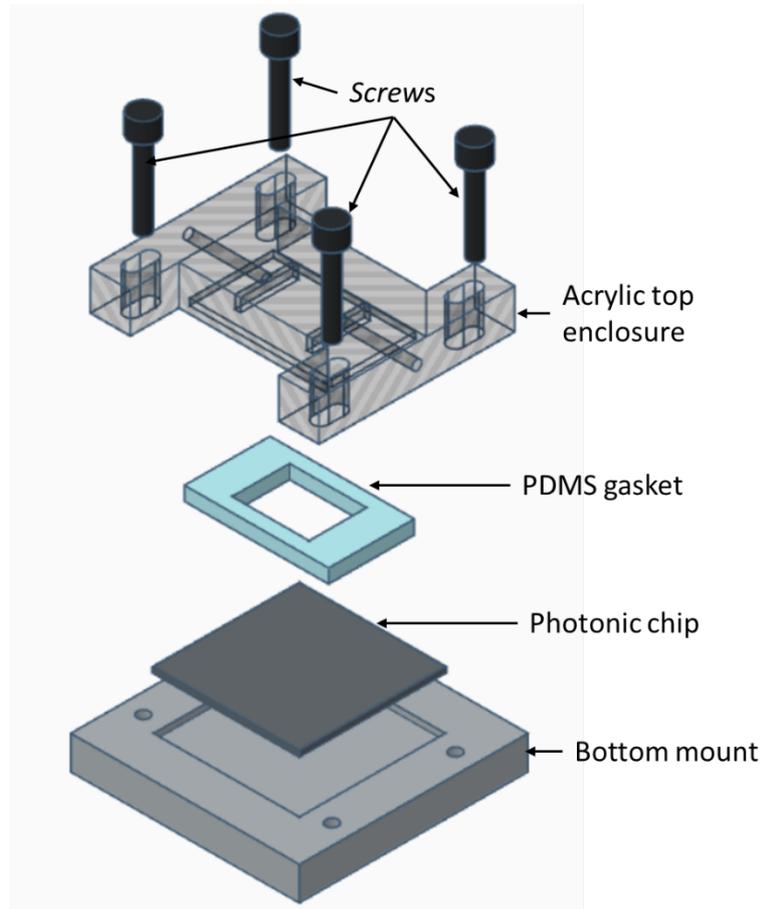

**Supplementary Fig. 1.** Exploded view representation of the Pb$^{2+}$ ion sensor assembly, consisting of the bottom mount, photonic chip, PDMS gasket, acrylic top enclosure, and the screws, which are used to secure the assembly, and prevent microfluidic chamber leakage. The assembly is mounted on top of a TEC controller for thermal stabilization.

**Supplementary Note 2: Definition of FoM for the Optimization of Surface Sensitivity**

The cross section of a slot waveguide is illustrated in Supplementary Fig. 2a, with strip and slot width indicated; 20 nm of SiO$_2$ is deposited on the strip via ALD prior functionalization (Fig. 3a of the main text). For instance, the electric field distribution of the implemented slot waveguide with strip and slot width of 240 nm is shown in Supplementary Fig. 2b. Higher surface sensitivity implies that power of the optical mode about the surface of the strip width will be higher. To that effect, the FoM which considers the confinement factor 10 nm from the waveguide surface, with 20 nm of SiO$_2$ cladding, is derived from the following:

$$FoM = \frac{\partial n_{eff.}}{\partial n_{confine}} = \frac{n_g}{n_{confine}} \frac{\int_{confine} \varepsilon |\vec{E}|^2 dA}{\int \varepsilon |\vec{E}|^2 dA} \quad (2.1)$$

n$_{eff.}$ and n$_g$ are the effective index and group index of the optical mode respectively, $\vec{E}$ is the electric field of the waveguide mode, n$_{confine}$ and $\varepsilon$ are the material index and the dieletric permittivity of the medium. In the computation, the medium surrounding the waveguide is taken to be water. The 2D integral in the numerator is to be taken about the 10 nm region which surrounds the waveguide and its cladding, while the integral at the denominator is considered about the entire space. The computed FoM of the slot waveguide is indicated in Fig. 2b of the main text. It is known that when the slot width is comparable to the exponential decay length of the fundamental eigenmode, optical power perpendicular to the high-index contrast interfaces is amplified [1–6]. As a comparison, the optical power 10 nm about the surface of strip TE and TM waveguides with 20 nm of SiO$_2$ cladding was also computed. The cross section of the strip waveguide is illustrated in Supplementary Fig. 2c. The FoM of the strip waveguides at the fundamental TE and TM polarization are computed as a function of strip widths in Supplementary Fig. 2d. On the 220 nm SOI platform, at λ = 1.55 μm, the optimal waveguide width for surface sensing at the TE polarization is found to be 270 nm. The FoM for TM strip waveguides is found to be lower than that of TE. This is because for TM polarization, the optical field is intensified at the top and bottom interfaces of the waveguide (Supplementary Fig. 2f)[7], unlike TE (Supplementary Fig. 2e), which is lateral [7]. The field at the bottom interface of the TM waveguide does not contribute to surface sensitivity. Furthermore, as the BOX has higher material refractive index than DI water, the field intensity at the bottom interface of the TM strip waveguide will be higher than that at the top (Supplementary Fig. 2f). From Supplementary Fig. 2d it can be seen that the FoM of the TM strip waveguide plateaus when strip width is larger 400 nm; the cross sectional electric field distribution of a TM strip waveguide with width of 425 nm is indicated in Supplementary Fig. 2f.Comparing the FoM of the slot waveguides (Fig. 2b of the main text) against that of the TE and TM slot waveguides, it can be seen that significantly higher surface sensitivity can be realized for slot waveguides. To that effect, slot waveguides are implemented for the $Pb^{2+}$ photonic sensor presented in this work.

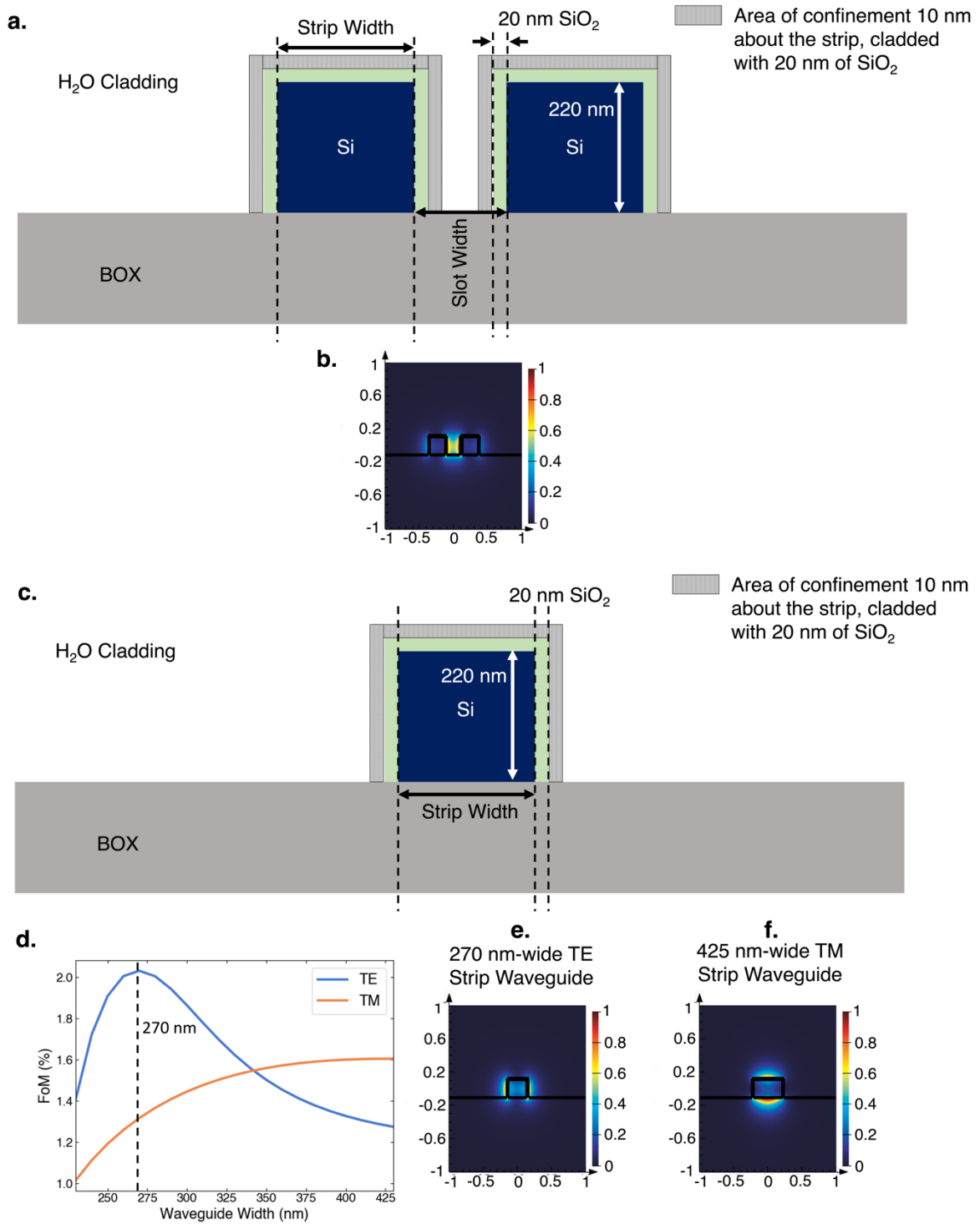

**Supplementary Fig. 2. Definition of FoM for the optimization of waveguide surface sensitivity. The height of the waveguide layer is 220 nm with a 20 nm SiO₂ layer surrounding the waveguide, deposited via ALD. a,** Cross-sectional schematic of the slot waveguide with definition of slot and strip width illustrated; the area of confinement is illustrated. **b,** An instance of electric field distribution pertaining to a slot waveguide with slot and strip width of 240 nm. **c,** Cross-sectional schematic of the strip waveguide(TE/TM) with definition of strip width

illustrated; the area of confinement is illustrated. **d,** Sensor surface sensing FoM for strip waveguides (TE and TM) as a function of strip width. Instances of electric field distribution pertaining to **e,** TE strip waveguide with width of 270 nm, and **f,** TM strip waveguide with width of 425 nm.

**Supplementary Note 3: Derivation of Equation 3 in the main text, relating designed loss (water absorption) with splitter power ratios**

Condition 1 is illustrated in Supplementary Fig. 3a. A lightwave ($I = |E^2|$) is injected into an asymmetrical splitters, with two power splitting ratios; $S_1, S_1'$ (input splitter) and $S_2, S_2'$ (output splitter) where, $S_1 = S_2$ and $S_1' = S_2'$, $S_{1/2}' \neq 0.5$. Assuming the splitters are lossless, energy conservation dictates that $S_{1/2} = 1 - S_{1/2}'$. After the input splitter, the field is separated into the sensing and reference arm, yielding $E_1$ and $E_2$ respectively. These two terms are related to the electric field at the input of the splitter ($E$) through the splitting ratio such that $E_1 = E\sqrt{S_1}$ and $E_2 = E\sqrt{S_1'}$. As the light propagates through the sensing arm, it will accumulate attenuation from water absorption which we define as $e^{-\alpha L}$, with L being the sensing arm length and α is the attenuation constant, as compared to the reference arm. We can express the phase difference between the two MZI arms as $e^{-j\Delta\phi}$.

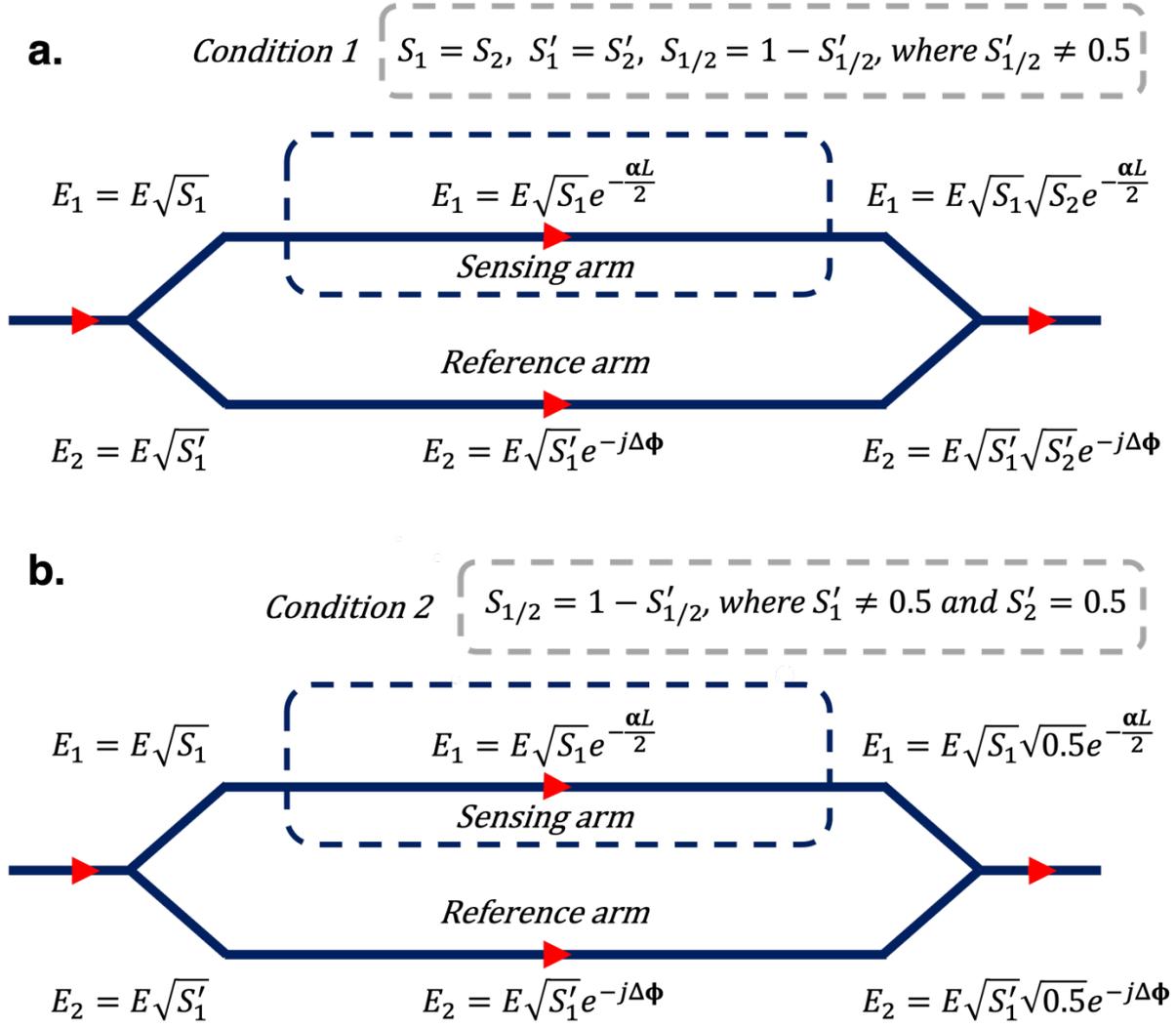

**Supplementary Fig 3. Architectures of two MZI-based sensor designs with a,** Condition 1, $S_1 = S_2, S_1' = S_2', S_{1/2} = 1 - S_{1/2}'$, where $S_{1/2}' \neq 0.5$, and **b,** Condition 2, $S_{1/2} = 1 - S_{1/2}'$, where $S_1' \neq 0.5$, and $S_2' = 0.5$.

According to Fig. 2c of the main text, it can be seen that the architecture in Supplementary Fig. 2a, condition 1 imposes lower requirements on splitter asymmetry as compared to Supplementary Fig. 2b, condition 2.

Lastly, as the two fields recombine at the output splitter, the total output intensity will be of the form:

$$I_{out} = E^2 |\sqrt{S_1}\sqrt{S_2}e^{-\frac{\alpha L}{2}} + \sqrt{S_1'}\sqrt{S_2'}e^{-j\Delta\phi}|^2 \quad (3.1)$$

In order to maximize sensor visibility, $I_{out} = 0$ during destructive interference condition. As a result, (S2.1) can be reduced to the following form,

$$\sqrt{S_1}\sqrt{S_2}e^{-\frac{\alpha L}{2}} = \sqrt{S_1'}\sqrt{S_2'} \quad (3.2)$$

An alternative MZI architecture (condition 2) is also illustrated in Supplementary Fig. 3b. In this design, an arbitrary and 3-dB splitter are used at the input and output respectively; $S_1' \neq 0.5, S_2' = 0.5$. The output of the MZI can be defined as the following.

$$I_{out} = E^2|\sqrt{S_1}\sqrt{0.5}e^{-\frac{\alpha L}{2}} + \sqrt{S_1'}\sqrt{0.5}e^{-j\Delta\phi}|^2 \quad (3.3)$$

Similarly, to maximize sensor visibility, we set $I_{out} = 0$. As such, Equation 3.3 can be reduced to.

$$\sqrt{S_1}e^{-\frac{\alpha L}{2}} = \sqrt{S_1'} \quad (3.4)$$

Via Equation 3.2 and 3.4, we determine the splitting ratio of the arbitrary splitters as a function of designed losses for the MZI architectures in Supplementary Fig. 3a-b; designed loss from water absorption is assumed to be the only source of optical loss. It can be seen that the MZI structure illustrated in Supplementary Fig. 3a reduces the asymmetrical requirement in power splitting which significantly alleviates requirement on fabrication ratios; accurate fabrication of highly asymmetrical power splitters is challenging; small variations in splitter dimensions will result in significant changes from the intended design.

**Supplementary Note 4: Detection of $Sn^{2+}$ through EDX analysis after functionalization**

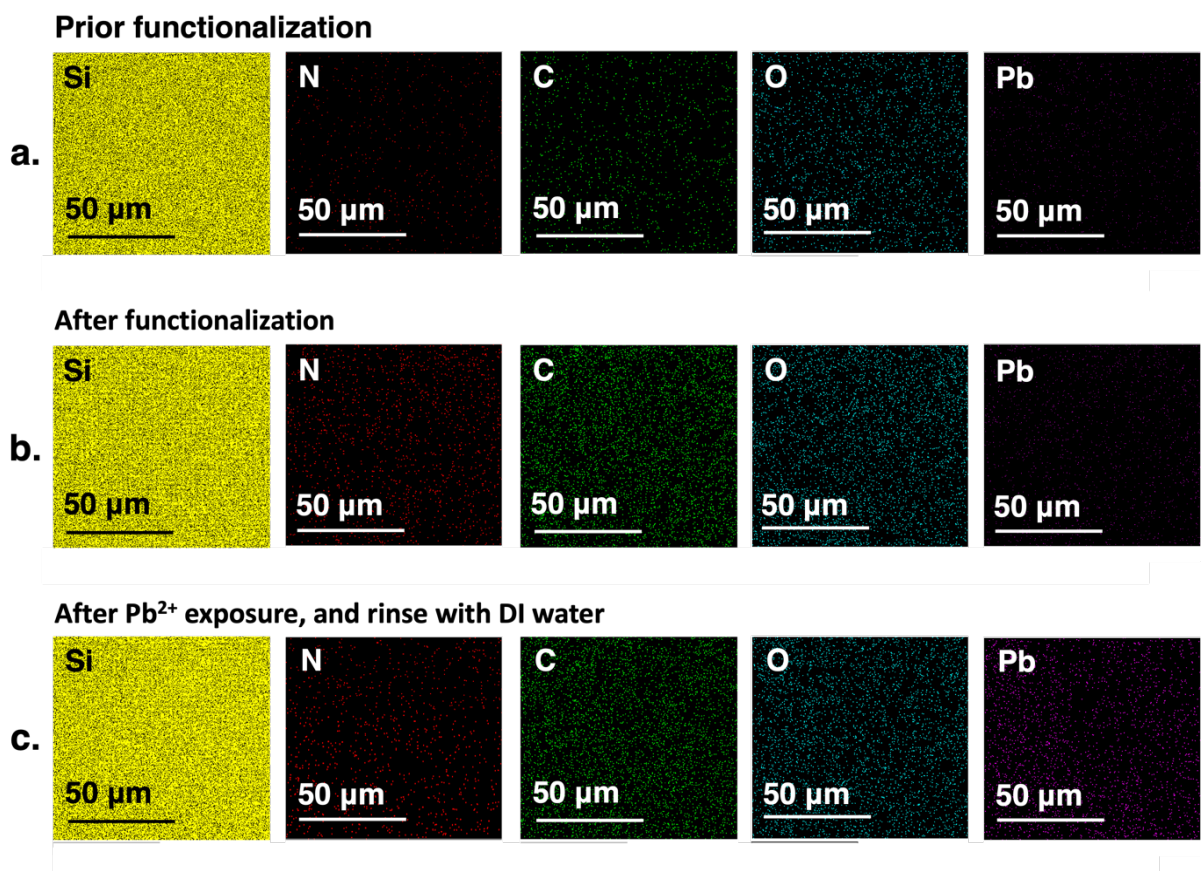

**Supplementary Fig. 4 EDX analysis of the photonic chips pertaining to Si, N, C, O, and Pb elemental composition a,** prior functionalization, **b,** after functionalization, and **c,** after $Pb^{2+}$ exposure and rinse with DI water and dry ($N_2$).

In addition to the XPS results shown in the Fig. 3b-d, and Fig. 4I of the main text, an EDX analysis (Supplementary Fig. 4) was carried out to evaluate the elemental composition of Si, N, C, O and Pb. The analysis of N, C, and O indicates the viability of the Fischer esterification protocol that was developed in this work[8]. By comparing Supplementary Fig. 4a with Supplementary Fig. 4b or Supplementary Fig. 4c, an increase in N and C elemental composition can be observed. This is in line with the conclusion derived from the N 1S and C 1S regions of the XPS spectrums shown in Fig. 3b-c of the main text. We were not able to observe a change in O elemental composition in the EDX analysis. This is due to the fact that the O signal is attributed from the functional layer, as well as the 20 nm $SiO_2$ that was deposited on the slot waveguides prior functionalization. As the 20 nm $SiO_2$ layer is the primary contributor to the O signal in Supplementary Fig. 4[9], a clear change in signal intensity cannot be seen. To further validate the capacity of the functional layer to bind with $Pb^{2+}$ ions, the functional layer is exposed to $Pb^{2+}$ for 120 s, followed by flushing with DI water and drying ($N_2$). By comparing the rightmost column of Supplementary Fig. 4b and c, the appearance of Pb via the binding capabilities of the functional layer can be clearly observed.

**Supplementary Note 5: Detection of $Sn^{2+}$ through XPS analysis after functionalization**

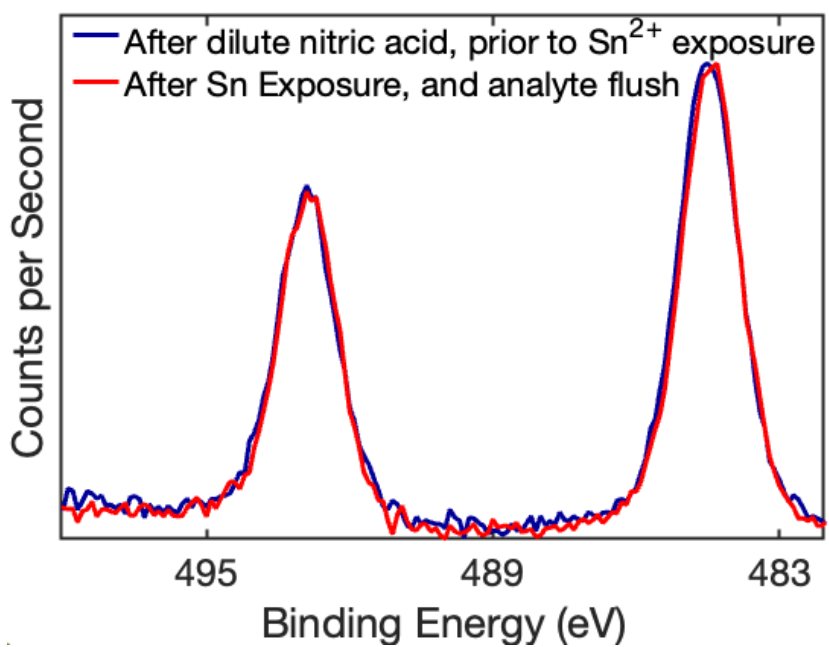

**Supplementary Fig. 5** XPS spectrum measured after dilute nitric acid purification and prior to $Sn^{2+}$ exposure, and after Sn exposure with DI water rinse and dry ($N_2$).

Due to the application of $H_2O$ as a green solvent for the developed reaction, Brønsted acid catalyst such as $H_2SO_4$ is incompatible, in view of its drastic decrease in catalytic activity, in the presence of $H_2O$[10]. To that effect, the Lewis acid catalyst, $SnCl_2$ is utilized, which has a catalytic activity that is more resilient to the presence of $H_2O$[11,12]. While Fischer esterification is favored when $H_2O$ is removed as the reaction proceeds (dehydrative esterification), for the reaction disclosed in this work, Sn is embedded within the $SiO_2$ substrate, forming a heterogeneous catalyst. Thereby, improved catalytic activity[13] that favors esterification in the presence of $H_2O$ is achieved[14]. Evidence of successful Fischer esterification is indicated by the XPS N 1S, C 1S, and O 1S data in Fig. 3b-d of the main text respectively, and the EDX analysis in Fig S4 (See Section S4 in Supporting Information). Furthermore, Supplementary Fig. 5 shows the XPS results of the functional layer after dilute nitric acid purification (process in Fig. 3a of the main text), prior to $Sn^{2+}$ exposure. It can be seen that presence of Sn cannot be eliminated via the purification step. We note that heterogeneous catalyst displays improved catalytic activity that favors esterification, even in the presence of $H_2O$. Supplementary Fig. 5 also shows the XPS measurement of the functional layer after $Sn^{2+}$ exposure, followed by DI water flush and drying. As the DBTA crown ethers that undergoes amine conjugation subsequently do not bind to $Sn^{2+}$ ions, it can be seen that the XPS spectrum is similar to that prior $Sn^{2+}$ exposure. The subtraction of the narrow scan XPS spectrum before ion interaction was subtracted to that after ion interaction, referring to the normalized XPS spectrum as shown in Fig. 5 of the main text.

## Supplementary Note 6: Measured optical spectrum of the photonic sensor when the sensing region is exposed to air

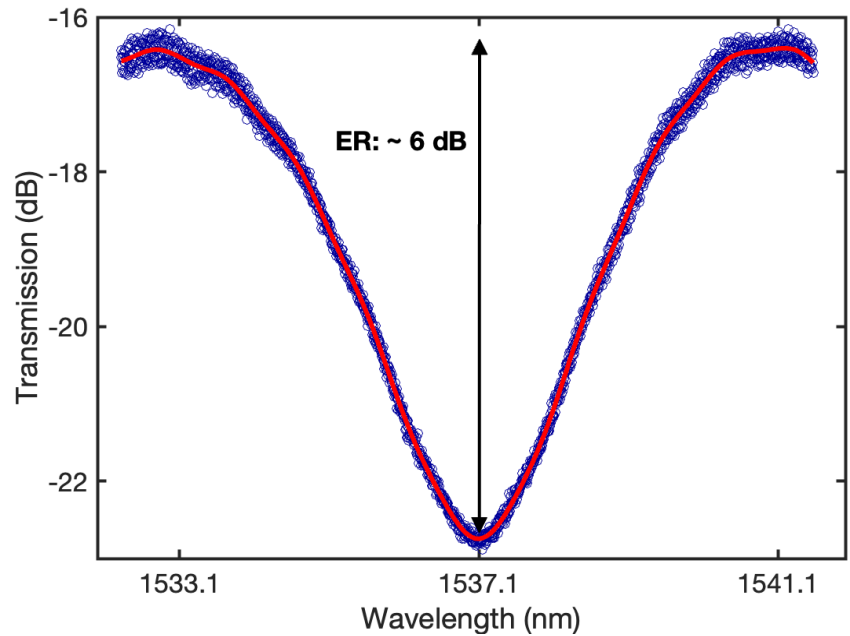

**Supplementary Fig. 6. Measured optical spectrum of the photonic sensor when the sensing region of the photonic sensor is exposed to air.** Poor visibility, as implied from the interferometric spectrum ER is observed, which is resultant when the designed losses are non-optimal to the asymmetrical splitting ratios.

## Supplementary Note 7: Details on chip fabrication

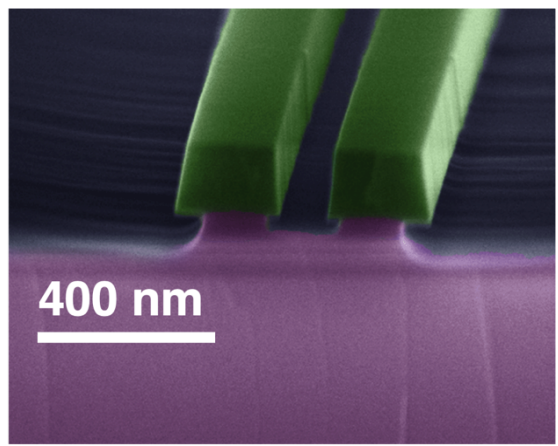

**Supplementary Fig. 7** The SEM image (false colour) of fabricated slot waveguide, with strip and slot widths of 240 nm.

The fabrication of the sensor chips starts from commercially available 200 mm silicon on insulator wafers with 3 μm thick buried oxide and 220 nm thick device layer. First, the wafers are cleaned using a heated acetone bath kept at 55°C and rinsed in methanol, isopropanol and DI water. Then, an adhesion promoter (Surpass 4000) and electron-sensitive resist (ma-N 2403) are spin-coated onto the wafer and baked for 2 minutes at 90°C. The thickness of the E-beam resist after spin coating is ~300 nm (achieved at spin speeds of 3000 rounds per minute). A discharging layer (Espacer-300Z from Showa Denko Inc.) is applied to minimize charging effects. The wafer is patterned using E-beam lithography (ELS-HS50 from STS-Elionix) with a 50 kV accelerating voltage and a beam current of 5 nA. After development in RD6 (Futurexx Inc.) for 80s and rinsing in DI water, the waveguides are etched using ICP-RIE (RIE-230iP from Samco Inc.) with a gas chemisty of $CF_4$ and Ar at a pressure of 1 Pa, ICP Power of 300 W and 100 W of bias Power. The wafer is then ashed in $O_2$ plasma to strip away any remaining E-beam resist and to remove residual fluoropolymer formed during the etching process and thoroughly cleaned in Piranha solution, followed by a DI water rinse. Subsequently, the wafer was cladded with 2 μm of $SiO_2$ deposited at 350°C vie PECVD (Samco PD-220NL from Samco Inc). The wafer was then baked at 115 °C and silanized in an oven (TA Series from Yield Engineering Systems Inc.) to increase adhesion promotion of photoresist. Then, a thin (~1 μm) AZ 3312 photoresist layer is spin-coated and softbaked at 110°C for 60s. The sensing trench pattern is exposed into the resist using a maskless aligner (MLA-150 from Heidelberg Instruments Mikrotechnik GmbH) with a laser source centered at 405 nm. The resist was then post-exposure baked at 110 °C for 60 s and developed using AZ 726 MIF developer (Microchemicals GmbH) for 60 s. A diluted buffered oxide etchant solution was then used to open the sensing trenches, exposing the waveguides in the sensing arm. To avoid overetching, which could suspend the waveguides and generally change the structural cross-section from the intended design, the etching depth was monitored during the etching process using both profilometry (Dektak-XT from Bruker Corporation) and reflectometry (F50-UVX from Filmetrics Inc.). Once a total etch depth of 2 μm is reached, the resist was stripped in oxygen plasma (e3511 wafer asher from ESI Inc.), followed by thorough cleaning in acetone, isopropanol and DI water to prepare the wafer for the crown ether functionalization step.

**Supplementary Note 8 $Pb^{2+}$ ion photonic sensor selectivity test against $Na^+$, $Mg^+$, $Cu^+$, $Sn^{+2}$, $Ca^{+2}$, $Fe^{2+}$**

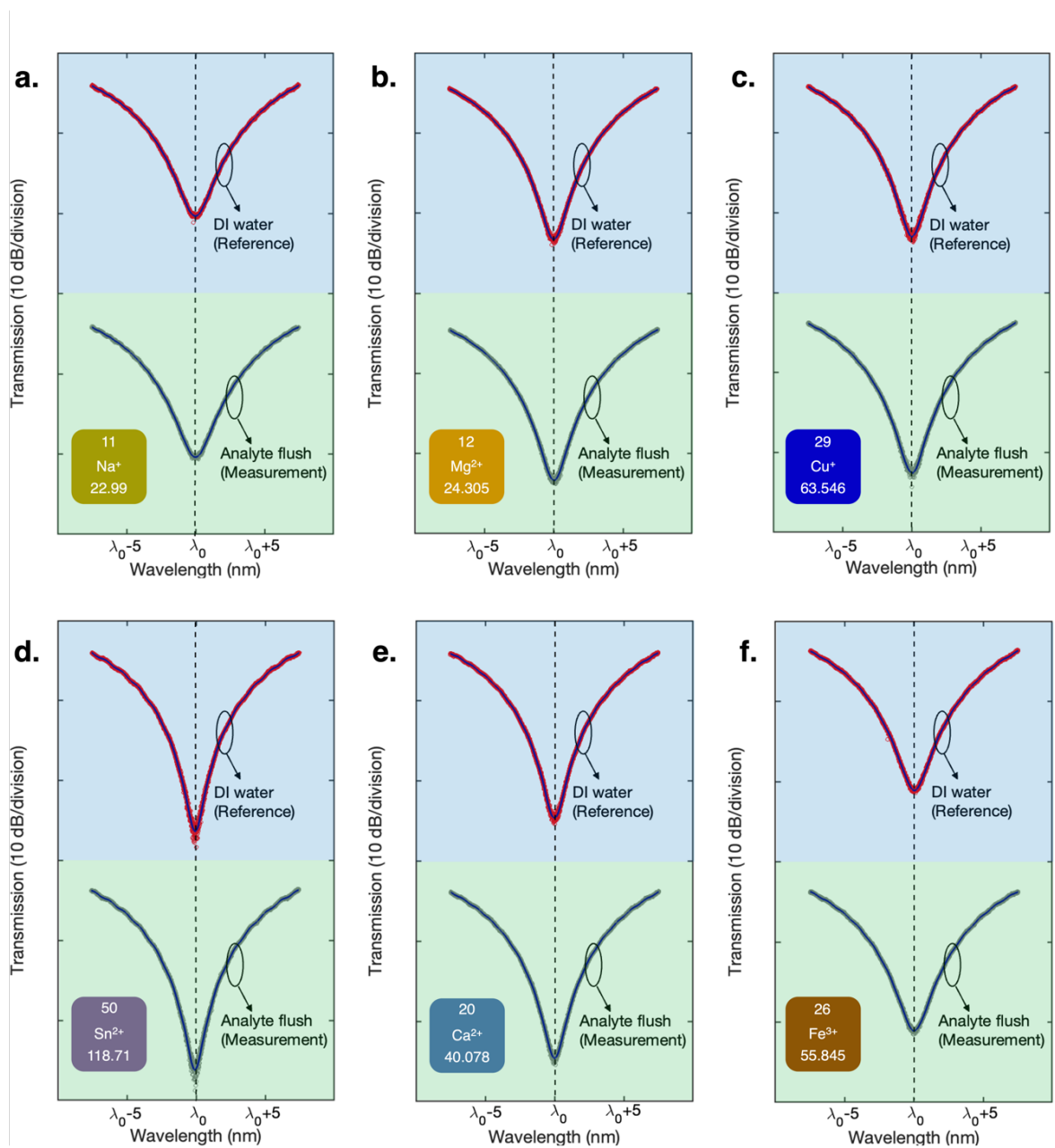
43

**Supplementary Fig. 8 Selectivity performance of the Pb$^{+2}$ ion sensor against a,** Na$^+$, **b,** Mg$^+$, **c,** Cu$^+$, **d,** Sn$^{+2}$, **e,** Ca$^{+2}$, **f,** Fe$^{+2}$ at 15 ppb where no shifts in the interferometric spectra indicative of ion binding is observed across all the tested ions.